\documentclass[10pt]{article}
\usepackage{graphicx}
\begin{document}

\centerline{\bf Sexual reproduction from the male (men) point of view}
\bigskip

\centerline{D. Stauffer$^*$ and S. Cebrat}

\bigskip
\noindent
Department of Genomics, 
Wroc{\l}aw University, ul. Przybyszewskiego 63/77, 51-148 Wroc{\l}aw, Poland

\medskip
\noindent
$^*$ Visiting from Institute for Theoretical Physics, Cologne University, 
D-50923 K\"oln, Euroland

\bigskip
Abstract: To counterbalance the views presented here by Suzana Moss de Oliveira,
we explain here the truth: How men are oppressed by Mother Nature, who may have 
made an error inventing us, and by living women, who could get rid of most of 
us. Why do women live longer than us? Why is the Y chromosome for men so small?
What are the dangers of marital fidelity? In an appendix we mention the 
demographic challenges of the future with many old and few young people.

\bigskip

\bigskip

\section{Introduction}
The French revolution was based on Freedom, Equality, Brotherhood and lead
to the first declarations of ``Droits d'Homme''. Now, unfortunately, we are 
oppressed by Sisterhood, which claims human rights also for women. With this
oppression of men by women and of the present authors by the author of the 
accompanying chapter in this sociophysics school, we ask why do we exist
at all? Has Mother Nature made an error when inventing sexual reproduction
in addition to the older and simpler cloning (most of the time) of asexual 
bacteria, bdelloid rotifers, or other asexual species? As the Gena Rowlands
in the Kirk Douglas movie ``Lonely are the Brave'' pointed out, she would
not have anything to do with any male if they were not needed to produce 
babies. 

When we deal with sexual reproduction we have in mind species with a 
separation of male and female individuals where only the females give birth; 
and asexual reproduction refers to pure cloning, without any bacterial
``parasex'' (exchange of DNA). Thus sexual reproduction has a disadvantage 
by a factor of two over asexual cloning, since males do not get pregnant
(with few exceptions like Arnold Schwarzenegger in ``Junior''.) 

\section{Extrinsic reasons}
Our genetic properties are stored in the DNA of the genome, and are given on to 
our children. The same holds for bacteria and other forms of life. During 
the copying of the DNA some errors may occur, which are called mutations 
in biology and usually make life more difficult in the form of hereditary 
diseases. If the whole genome is stored only once one calls it haploid;
if it is stored twice, in two sets of chromosomes, it is called diploid.  

Apart from the mutations, in asexual reproduction the offspring has the
same genome as the parent; for example if one bacterium splits, each of the 
two new bacteria has the same genome. For sexual reproduction, the diploid
offspring has a mixture of the genomes from father and mother, who had produce 
haploid gametes (sperm cells and ovum) which were fused into a diploid
zygote, growing into an embryo. Thus each 
child is different from its parents and its siblings (identical twins excepted).
Sexual reproduction thus produces more diversity that asexual one. 

This diversity is a disadvantage if biological evolution optimises the genome
for a fixed environment, through selection of the fittest. Once the optimum was 
found one should stay with it. But physicists \cite{schneider} know from trying 
to find the 
ground state of a spin glass or other frustrated structure, one often ends 
up in one of the many local optima and not the single global optimum. Thus
to find a better local optimum, or even the elusive global one, diversity
can be good. In the simulated annealing method physicists have used for 
such optimization, the positive temperature produced this diversity away
from the local optimum. In nature, the mutations as well as the greater
diversity from sexual reproduction have caused the evolution of life from
simple bacteria to its pinnacle, the Herr Professor.  

Also, the environment is not fixed. Temperatures have changed in the past, and
if the temperature drops as in the movie ``The Day after Tomorrow'',
then the Inuit (=Eskimo) will simple move from Northern Canada to Florida
and hunt whales from the ice swimming off the coast of Miami Beach. The
greater human diversity thus allows some humans to survive this catastrophe.
Had mankind been transformed into a single race of Hitler's Aryan Herrenmenschen
they would not have been accustomed to the cold and all died \cite{cebpek}.

More explicitly for sexual versus asexual reproduction, \cite{whysex} used
the Penna ageing model (see our Appendix 1 or our books \cite{books}) to 
compare the response of a population to a sudden catastrophe, simulated by 
assuming that a previously good version of a gene suddenly threatens life
(like adjustment to warm climate threatens life after a sudden drop in 
global temperature). The asexual version was superior to the sexual one
before the catastrophe but died out after it. In contrast, the sexual 
version was able to recover. 

Another advantage of sex are parasites. They have to adjust to their host,
which is easier for the asexual case with little changes of the host with 
time, than for the sexual case where the host changes genetically from one 
generation to the next. Computer simulations \cite{howard}, also for the
Penna model \cite{martins}, confirmed the advantage of sexual reproduction
quantitatively.

Thus we have good external reasons to justify the existence of sexual 
reproduction in real nature, due to the increased diversity.

\section{Intrinsic reasons}

Far less clear is the justification of sex for purely intrinsic reasons,
without catastrophes, parasites, ... A simulation by zoologist Redfield
\cite{redfield} triggered many physics articles like \cite{zorzenon,anais}. 
If mutations happen as seldomly for the male as for the female, and if
one distinguishes between the rare dominant and the more widespread recessive 
mutations, then asexual cloning is worse than sexual reproduction in spite 
of having twice as many births per individual. With a male mutation rate 
much higher than the asexual or female mutation rate, sex looses against 
cloning. (Mutations are dominant if they affect us even if stored in only
one of the two sets of the diploid genome, while recessive mutations affect
us only if present in both sets. For asexual individuals with haploid genomes
each mutation is dominant.) 

However, the Redfield model, similar to Weidlich's methods in sociophysics
\cite{weidlich}, is not a proper agent-based model and treats probability
distributions instead of individuals. It also does not include the ageing
of individuals, an effect well known to the present authors. Thus simulations
of the Penna bit-string model \cite{penna} (see appendix) are more appropriate. 
Here each individual is simulated in many stages from birth to death.

\begin{figure}[hbt]
\begin{center}
\includegraphics[angle=-90,scale=0.45]{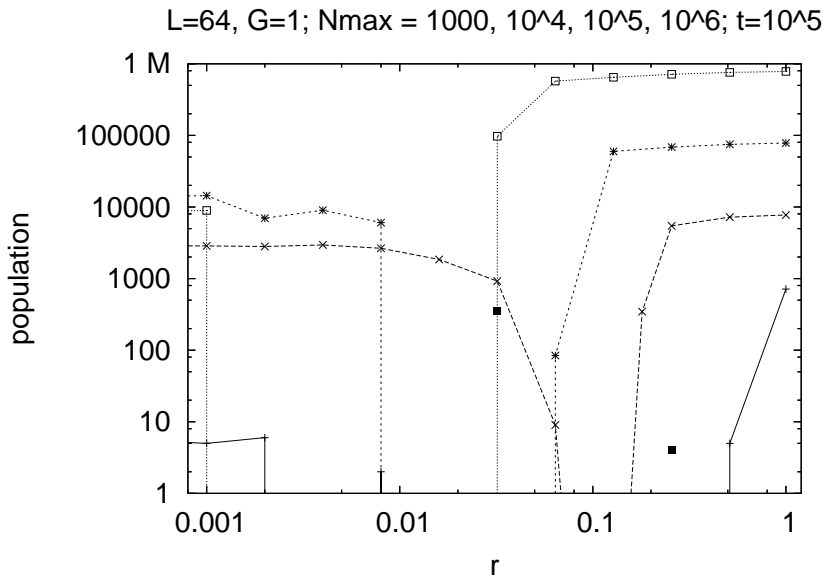}
\end{center}
\caption{Sexual Penna model; populations versus recombination (= crossover) 
rate $r$ for various values of the carrying capacity $N_{\max}$. The gap at 
intermediate $r$ shifts to smaller $r$ for increasing $N_{max}$. To the left 
of the gap we have complementarity with many mutated bits; to the right 
we find Darwinian purification selection with much less bits mutated.
}
\end{figure}

\begin{figure}[hbt]
\begin{center}
\includegraphics[angle=-90,scale=0.45]{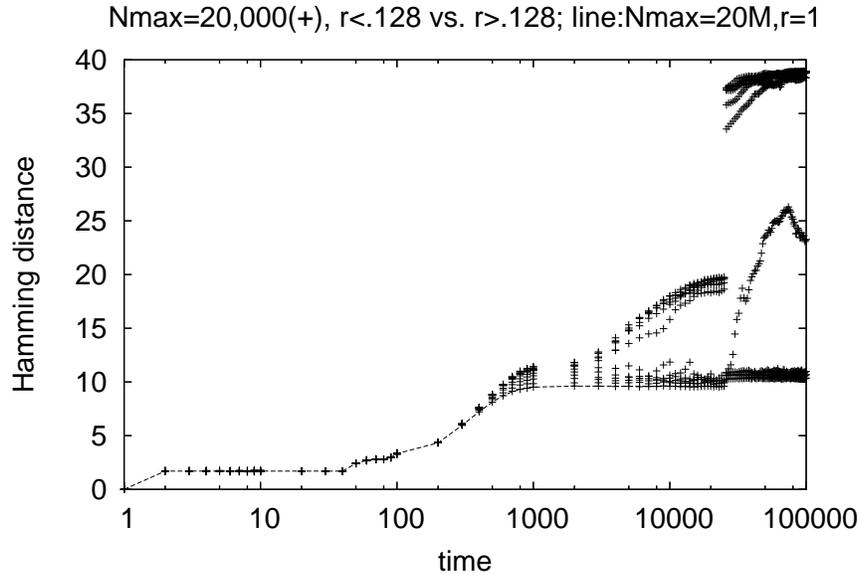}
\end{center}
\caption{Illustration of complementarity and gamete recognition. For
small recombination rates $0 \le r \le 0.016$, the average Hamming distances 
approach 20, and then move close to 40 after genome recognition is switched
on at $t = 25,000$: Complementarity with about half of the 40 bits mutated.
For $r = 0.032$ and 0.064 the population dies out, for $r=0.128$ it does
not know what it wants, and for larger $r$ relatively few bits are mutated:
purification independent of population size.
}
\end{figure}

\begin{figure}[hbt]
\begin{center}
\includegraphics[angle=-90,scale=0.45]{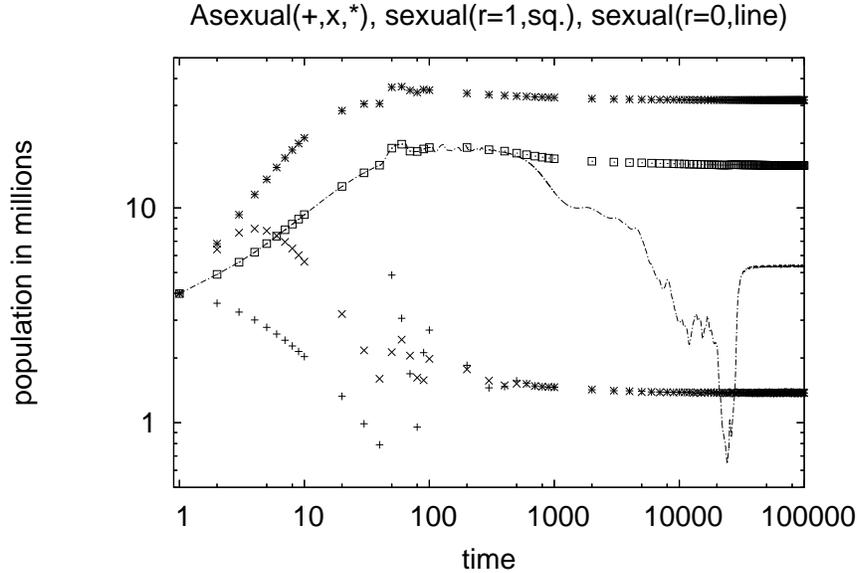}
\end{center}
\caption{Comparison of sexual with asexual reproduction. The sexual case (line)
is simulated with gamete recognition, for both $r=0$  and $r=1$,
while for the asexual haploid case (other symbols) there is no crossover, 
no complementarity and no gamete recognition. Some asexual cases are 
worse but one is better than the sexual strategy.
}
\end{figure}

Again, also in the Penna model no clear advantage of sex was found \cite{anais}
until Ref.\cite{martinsstauffer} assumed that mutations (= inherited diseases)
which finally kill us with certainty, already before reduce our health and 
increase our mortality. Moreover, F. Scharf (diploma thesis 2004, as
presented on page 91 in the second of Ref.\cite{books}) showed that preselection
of sperm cells, before fusion with the ovum, may give sex an advantage since
it is impossible for asexual cloning. For example, sperm cells with some 
genetic defect may swim slower than healthy ones and thus not reach the 
ovum in time.

Somewhat related is gamete recognition \cite{gamete}, where the ovum rejects
those sperm cells for fusion into a diploid zygote whose haploid genome is too 
similar to the haploid genome of the ovum. This effect is beneficial if the
population, due to a low recombination rate (see appendix), shows the recently
discovered genome complementarity \cite{zawierta,bonkowska,pekalski,pmco}.
In such a population nearly all individuals have the same bit-strings A and B
in their diploid genome, thus producing haploid gametes (ovum and sperm
cells) of type A only. An A sperm combined with an ovum of type A cannot 
survive  with many mutations, since then even recessive mutations affect our
health. The same happens with ovum and sperm cell both of type B. But if one
is of type A and one of type B, the A-B-zygote can survive even if half of the 
bits (alleles of the genome) are mutated, since there is always a one-bit 
combined with a zero-bit and thus for recessive mutations the health is not
affected. Thus high numbers of mutations can be tolerated in this sexual
version, while they lead to extinction in the asexual case. 

Figure 1 (after \cite{zawierta}) shows the two regimes of low and high 
recombination rates. Each curve has a gap in the middle where the population
dies out. For low $r$ the population survives with the help of the above 
complementarity trick; for high $r$ it survives through purification, i.e. the 
usual Darwinian selection of the fittest with a small number of bad mutations.
These data are obtained without the above gamete selection. If this 
gamete selection is
added to the model then the population size to the left of the gap (small
$r$) is strongly enhanced while the populations to the right of the gap 
($r$ closer to unity) barely change. 

Figure 2 illustrates through the Hamming distances this balance between 
complementarity at small $r$ and purification at large $r$, separated by 
extinction at intermediate $r$. These Hamming distances are the number 
of bits which differ in a position-by-position comparison of the two
bit-strings which form a zygote, taking into account the first 40 of the 
64 bits. For complementarity without gamete recognition, the whole diploid
population has two bit-strings A and B, each of which with about 20 bits zero 
and 20 bits one. The zygotes thus are of type AA and BB with Hamming distances 
0 and of type AB and BA with Hamming distances 40; the average Hamming distance
therefore is 20, as shown in Figure 2 near $t = 10,000$. The AA and BB will die 
out in the next iteration, the AB und BA will survive. After 25,000 iterations, 
gamete recognition is switched on, neither AA nor BB is allowed to form a 
zygote, and the Hamming distances approach 40, as shown in the interval
$26,000 \le t \le 100,000$. For large $r$ and purification, the number of 
mutated bits and thus the Hamming distance is much smaller, and the latter 
shows only a small jump from 9.6 to 10.3 (independent of $N_{\max}$)
when gamete recognition is switched on.

Population versus time is shown in 
Figure 3, with squares for $r=1$ and the line for $r=0$; in the latter case,
the population nearly dies out but at time = 25,000 iterations the gamete 
selection is switched on and the population is saved.

How does this sexual reproduction fare compare with the asexual one? Figure
3 shows that victory depend on details. In the lowest curve (+) we start with
only babies; thus at first the population goes down since nobody has reached
maturity yet. The second curve (x) starts with a random age distribution
and thus first increases. After 1000 iterations it has merged with the lowest 
curve and its population stays constant even to $t = 10^6$ beyond this plot. 
Both cases have an equilibrium population below the two sexual curves, i.e. 
we have justified the existence of men. 

However, this comparison is unfair. In the sexual cases the Verhulst 
death probability (see appendix) was applied to the babies only, while for
the asexual case it applied to all ages, with a stronger reduction of the 
population. If in the asexual case we apply the Verhulst factor to the babies 
only, as we did for the sexual case, then we get the highest curve (stars),
and sex is worse than asexual cloning.

Thus since 13 years \cite{redfield} the story is the same: whether sex is 
good or bad depends on details. Moreover, we do not yet understand why 
diploid hermaphroditism is not much better than both. There is much to do
to justify our existence: ``Men of all nations unite, you have nothing to lose 
but your computer time.''

\section{The role of inbreeding}

As has been shown in Fig. 1, reproduction success depends on the interplay
between the intragenomic recombination rate (crossover frequency) and the
size of population. Below a specific crossover rate populations prefer to
complement haplotypes instead of to intensively eliminate defective alleles. In
Fig.4. we have shown how this critical crossover rate depends on the
population size (M. Zawierta, personal communication). In the range of two
decades there is a power law relation. Nevertheless, the data shown in the
plot were obtained in simulations of panmictic populations. In such
populations females look for and choose randomly a sexual partner from the
whole population. In Nature the process of choosing the partner is usually
nonrandom and, what is more important, it is spatially restricted.
Individuals are looking for partners in their neighbourhood. Thus, the effect
of the population size should be considered as an effect of the inbreeding,
rather. Inbreeding (coefficient) is a measure of genetic relations between
individuals. If the individuals live in small ``inbreeding'' groups, then the
inbreeding coefficient is high and there is a high probability that they
share some fragments of the same ancestral genome.

\begin{figure}[hbt]
\begin{center}
\includegraphics[angle=-90,scale=0.45]{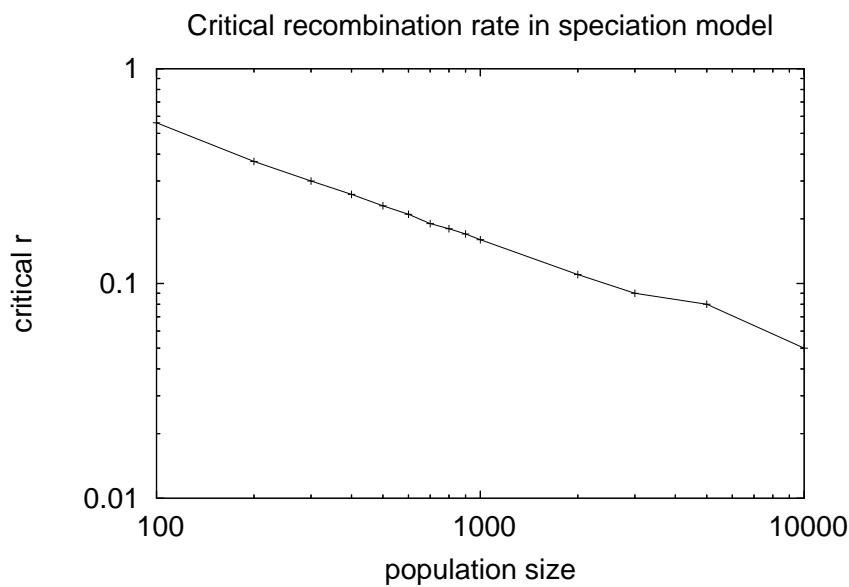}
\end{center}
\caption{Log-log plot of critical recombination rate $r$ versus population 
size. For $r$ below this value, complementary haplotypes (bit-strings) are
preferred. 
}
\end{figure}

\begin{figure}[hbt]
\begin{center}
\includegraphics[angle=-90,scale=0.45]{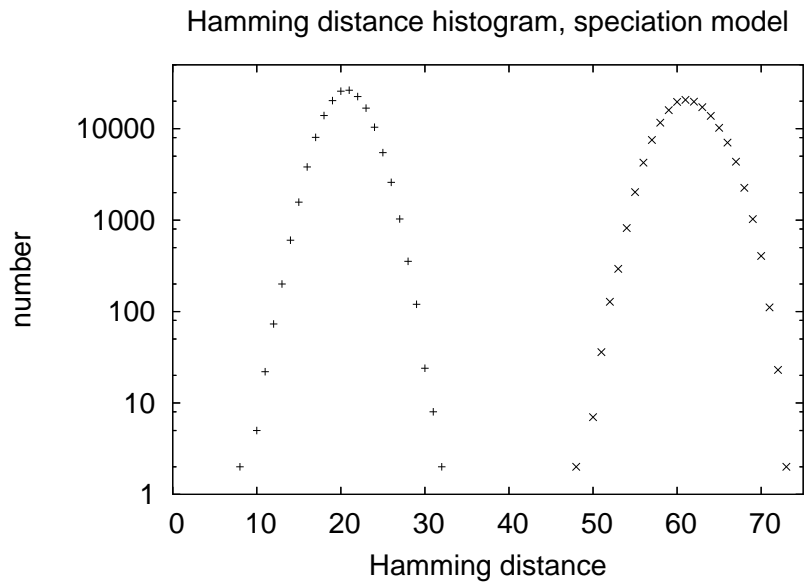}
\end{center}
\caption{Histogram for Hamming distances within one species (left peak) and 
between different species (right peak). The bit-strings had 128 bits.
}
\end{figure}

To study the effect of inbreeding, the simulation of evolution was performed
on lattices. The level of inbreeding was set by declaring the maximum
distance where individuals can look for partners and where they can place
their offspring \cite{zawierta}. The simulations were performed on a square
lattice $1000 \times 1000$. If the above distances were set to 5, the critical
crossover rate was around 0.2. Populations evolving under lower
recombination rate or shorter distances prefer the strategy of complementing
the haplotypes while under higher recombination rate or longer distances
they choose the strategy of purifying selection. Nevertheless, there are
very important consequences of such a kind of choice. The complementarity
evolves locally and remote subpopulations on the same lattice can have
different distribution of defective alleles in their haplotypes. Using some
tricks with coloring the individuals according to their genomes' structure
it has been shown that the lattice is occupied by individuals with different
genotypes but they are clustered. Individuals with the same genotypes
occupy the same territory (see http://www.smorfland.uni.wroc.pl/sympatry/
for some examples of simulations under different conditions). In Fig.5 we
have shown the Hamming distances between corresponding haplotypes (not
complementary, in the description on page 4 they correspond to pairs of
haplotypes AA and BB). The similarity of haplotypes inside a species is high
while between species it is low. Further studies have shown that for the
clustering only the central part of the genome is responsible. The lateral
part of the genome is much more polymorphic and decides on biodiversity,
rather than speciation. That is why the Hamming distances between homologous
haplotypes inside species are noticeable. These simulations show that
sympatric speciation is possible and there is no need for physical,
geographical or even biological barriers for the new species to emerge inside
the population of the older one.

\section{Why do women live longer than us?}

Of course, we die sooner because women oppress us. But since this truth
cannot be published in detail (just because it is true), we have to find 
other reasons.  The senile author is so thin and close to starvation because 
women eat his steaks and drink his beer. But since at least for rodents,
caloric restriction prolongs life (if one can call that life), this example
may not be convincing.

One genetic reason could be the difference between the two X chromosomes for
women, compared with one Y and one X chromosome for men. The Y chromosome, 
in comparison with the X chromosome, contains very little information. Thus
if a mutation creates an error in the single male X chromosome, the correct
information is lost. An error in one of the two female X chromosomes, on
the other hand, can still be balanced by the correct information on the 
other X chromosome. Quantitative simulation \cite{schneiderold} in a Penna
model with many chromosomes gave good agreement with reality: Male mortality
is about twice as high as female mortality, except that the two get relatively
close at very old age.

Mammals share this chromosome difference between male XY and female XX, while
for birds the situation is reversed: Same chromosomes for the males and 
different ones for the females. Thus the above argument would means that
male birds live longer than their female counterparts. The empirical
observations are contradictory, as reviewed in \cite{books,encybio}.

Male sleep is often disturbed, e.g. by a dean in faculty meetings, while their
wives can calmly deal with many children and household chores at home.
Does this stress kill us? Similarly, the stress of the industrial revolution
was held responsible for the mortality difference \cite{blickpunkt}. But
why do already in the first years of life the boys die more often than the 
girls? And why did Swedish men die sooner than Swedish women already 230 
years ago? Why are there important differences between different 
industrialized countries in the male-female difference of life 
expectancies? More literature, but not more answers, are found in 
\cite{encybio}. 

\begin{figure}[hbt]
\begin{center}
\includegraphics[angle=-90,scale=0.45]{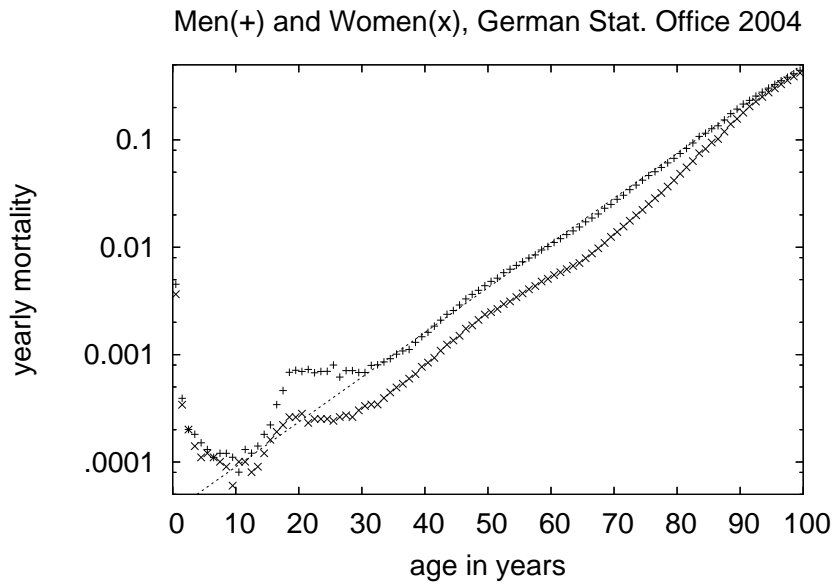}
\end{center}
\caption{Comparison of male and female mortality function $\mu$ for modern
Germany. The straight line indicates a Gompertz law \cite{encybio}. No
mortality deceleration is seen for adult men or women up to 100 years of age.
Data from 2003-2005 published in 2006. 
}
\end{figure}

Women should not be trusted anyhow, as the biblical story of Adam, Eve and the 
snake tells us. This is particularly true for mortalities, defined as the 
negative age derivative of the logarithm of the number of survivors at the
given age. Adult men obey nicely the Gompertz law of 1925 that the mortality 
$\mu$ increases exponentially with age, Fig.6. Women, in contrast, follow it
only for middle ages; in old age their mortality increases stronger with 
increasing age, until for centenarians the ratio of male to female $\mu$ 
is close to one, Fig.6. These lawless women then misled some into the belief
that there are strong downward deviations from the Gompertz law: ``mortality
deceleration''; just look at the $\times$ symbols in Fig.6 for ages 70 to 100. 
The true Gompertz region for women is 30 to 70 years, and a straight line 
extrapolated from there to older ages gives upward, not downward, deviations
from the Gompertz law. Only for ages beyond Fig.6 a mortality plateau
may appear \cite{robine}.

\section{Genetic marriage counseling}

1. One of the attributes of life is reproduction. Simulations of the 
age-structured populations using the Penna model show that if we set the upper
age limit for reproduction, we simultaneously set the maximum lifespan,
like in case of pacific salmon. Selection doesn't care about individuals
which cannot reproduce any more, and the genes necessary for them to survive
beyond the reproduction period accumulate mutations which kill the
individuals.

2. The above example corresponds to female menopause and a male condition
called sometimes andropause if both are set in the model at the same age.
Usually it is not true for real life. Women exercise menopause earlier than
men do andropause (if it is true that there is something like andropause).
In fact, men's reproduction ability does not stop in the middle ages and men
can reproduce late in their life. Menopause is a first-order phase transition
while men's reproductive ability slowly decays. 

It could be modeled just by setting the menopause (the upper age limit for
reproduction) only for women. In such a case the life expectancy for both
women and men stays the same as in the model version without menopause. Why?
Because men at the higher age can reproduce and their genes expressed after
the age of menopause of women are under selection, i.e. bad genes expressed
at old age spread in the population less easily than good genes. These correct 
genes could
be transferred to the daughter's genomes resulting in the higher life
expectancy of women. One can obviously conclude that women live longer due
to the cruel selection experienced by men. However, life is not so simple. Men
and women use to live in pairs and usually they swear to be faithful not
only until the menopause of the woman but to the end of their life. Thus,
when the wife reaches menopause the husband should simultaneously reach the
unreproductive age, too. In such a case we are back to the point 1: both
sexes have the same upper limit of reproduction period, given by female 
menopause, and they should die immediately after it.

There could be a few explanations of this inconsistency with reality:

\begin{itemize}

\item motherhood (child care) is necessary for higher chance of the offspring 
survival \cite{childcare},

\item  even a grandmother is necessary to increase the reproduction potential
of her children \cite{lahdenpera} (but see sec.3.5.3 in \cite{books}b).

\item older men are more attractive for young women when the latter ones are 
looking for partner (see e.g. the movie ``The First-Wives Club''),

\item  men have to be not faithful to their wife in menopause to secure the
longer lifespan of women (just read the newspapers).

\end{itemize}

One can notice, that there is no big difference between the last two items. If a
husband betrays his wife, then he should do this with younger women to succeed 
in the prolongation of the human lifespan for everybody. The only problem not 
solved until now is what should be the fraction of men unfaithful to their 
wives after menopause to produce this effect.

Men live shorter than women, Section 5. There are a lot of hypotheses (even more
than hypotheses) explaining this phenomenon \cite{books}. One of the 
explanations
is the role of men in cleaning out the genetic pool of the human population
of defective genes. The good and well-described example is the role of X
chromosome in this process. It is a rather large chromosome with high number
of genes. There are two copies of this chromosome in the woman's genome and
only one copy in the man's genome. Recessive defective alleles in one X
chromosome in the female genomes can be compensated by the corresponding
correct alleles while male genomes have no such a possibility. Each recessive 
defective X allele is seen for men like a dominant one and it is eliminated
by the purifying selection. Men should not complain about their situation
because Nature can be much more cruel. It produces male organisms which
clean up not only one chromosome but the whole genome, like drones in the
bees society. Drones are haploid and they have to be perfect to survive.
Sometimes drones are considered as extremely selfish individuals, nothing is
more wrong, they are exceptionally altruistic.

The stronger selection on genes located on X chromosomes in men's genomes leads to a significantly lower level of defective genes in the X chromosome when compared
with autosomes (all chromosomes except X and Y) \cite{Niewczas}. That is why the probability of disorders caused by defective alleles located on X chromosomes in
women is much lower than in men. As mentioned above, this purifying selection results in higher mortality of men what is observed in human populations as well as in
simulations \cite {schneiderold}, especially in the middle age \cite{Kurdziel}. In the standard Penna model a relatively low fraction of genes is expressed before the
reproduction age. In Nature, the fraction of genes expressed during development up to the puberty is probably much larger. Moreover, a large proportion of these genes
are expressed before the birth, what is usually not considered in the modeling. To reach the level of spontaneous abortion as in the natural human population (about
60 \%) \cite{Copp}, \cite{Hassold} a substantial fraction of genes have to be switched on before the birth. When such genes were introduced into the model, the other
effect emerged; among these genes some genes located on X chromosome were also expressed causing higher mortality of male embryos. Since the sex ration in human
populations at birth is close to 1, one should assume that for compensation the higher mortality of male embryos during the pregnancy, the sperm cells containing Y
chromosome should have about 50 \% higher probability of fertilizing the egg than sperm cells containing X chromosome \cite{Kurdziel}. This is a plausible hypothesis
because Y chromosome is smaller and the cells containing it could be faster.

On the other hand men should not complain about their role because they
themselves decided about that, putting the genes responsible for sex
differentiation into one copy of the sex chromosome. As \cite{onody} has shown,
this chromosome (known as Y chromosome) has to shrink during the process of
the genome evolution. \cite{onody} proposes a very plausible explanation of this
process of shrinking noting the role of selection which acts on different
number of sex chromosomes; 3 X chromosomes for each one Y chromosome in the
genetic pool. Additionally X chromosomes can recombine in the female genomes
while a Y chromosome never has a partner for recombination. Nevertheless,
there could be another explanation; men's genomes are under much weaker
selection pressure than female genomes because the reproduction potential of
the population depends on the fraction of female individuals. Women are 
promiscuous
and they can seduce men which already have children with other women. Thus,
to give birth the woman is indispensable but a large fraction of men is
dispensable (see the end of the Kubrick movie ``Dr. Strangelove''). And in fact,
these fractions of men which really have no
children are not under the selection. All chromosomes exercise alternatively
the selection in the male/female genomes with one exception --- Y chromosomes
evolving in the mostly dispensable men bodies, Fig.7.

To prove that the women's promiscuity is responsible for shortening the
men's life, some additional changes have been introduced into the Penna model
\cite{Biecek}. It has been assumed that man is indispensable for baby
survival and he has to stay only with one woman during the period of her
pregnancy. It means that the reproduction potential of population depends
equally on both, the fraction of females and males in the population, though
they need not swear to be faithful for life. In such a case the shrinking
effect of Y chromosome disappears, Fig.8.

Geneticists call the males a heterogametic sex because they produce two
different gametes (with X and Y). In all mammals males are heterogametic but
in birds females are heterogametic. To underline the difference in sex
determination geneticists call for birds the sex chromosome of males ZZ and of
females
ZW.  In the species with such a type of sex determination the shrinking of W
chromosome is not observed in the nature neither in the model. Moreover, it
seems that in such species the faithful males are losers.

Since not all men are necessary to keep the reproduction potential of the
population at high level, some of them, especially younger men with their
genomes not tested yet by life, can be altruistic and sacrifice their lives
\cite{klotz} for other members of the society. That is why we have armies and 
it is not wise to enroll women into the army because such a procedure diminishes
the reproduction potential of the population.

\begin{figure}[hbt!]
\begin{center}
\includegraphics[angle=-90,scale=0.35]{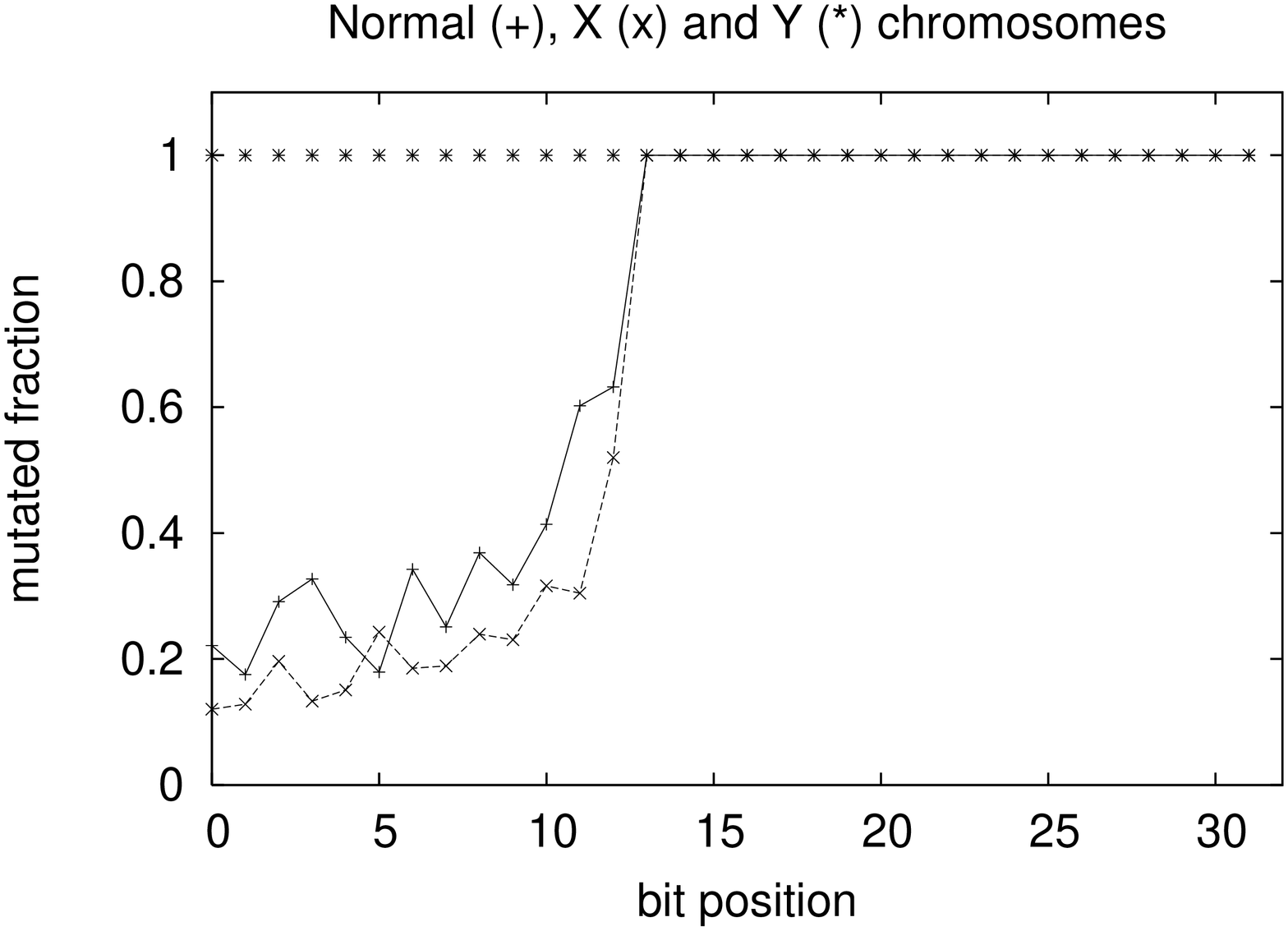}
\includegraphics[angle=-90,scale=0.35]{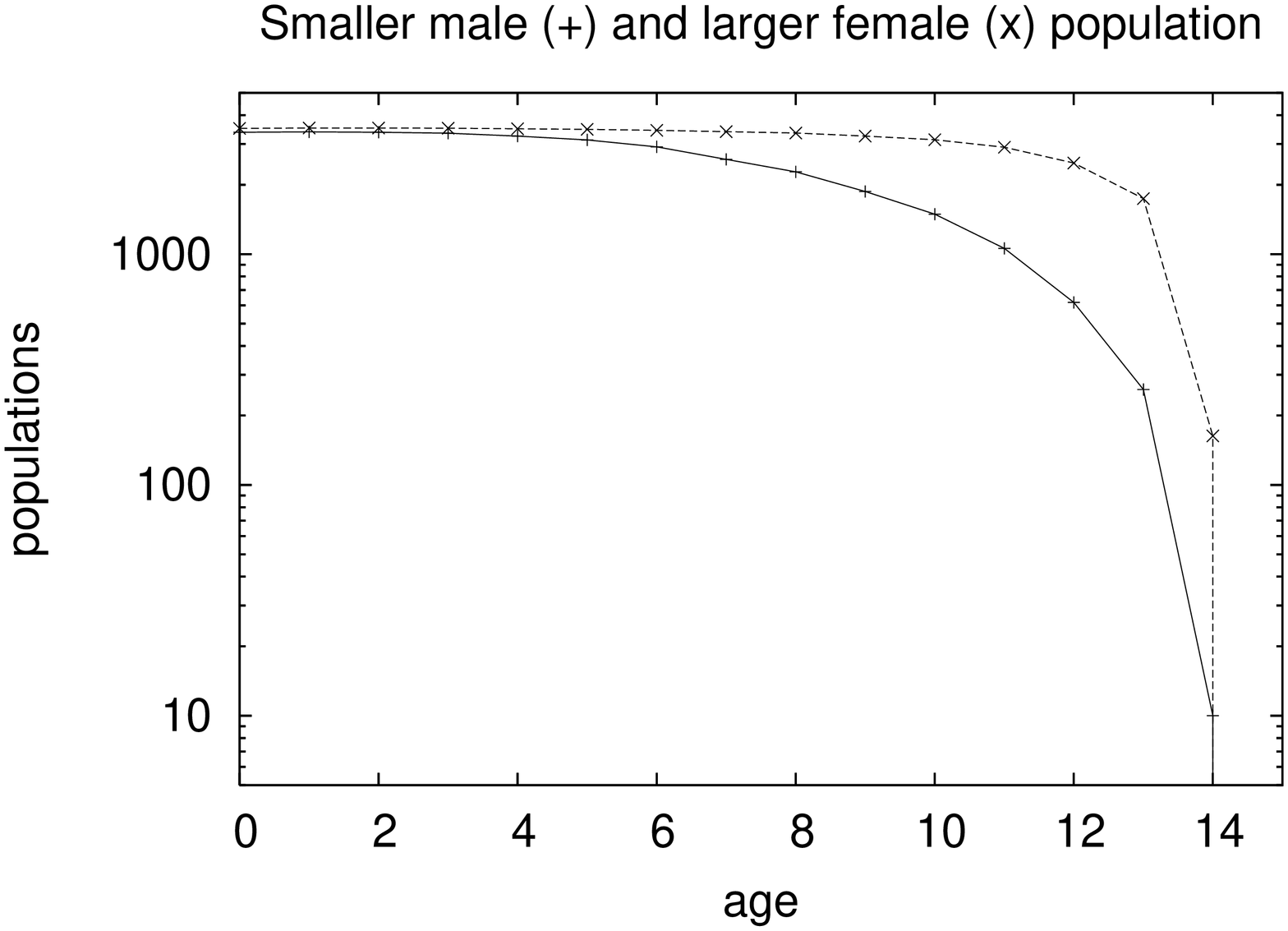}
\end{center}
\caption{
Top part: Fraction of defective genes in the autosomes and the two sex 
chromosomes if woman can seduce a man who already have been chosen by other 
woman as a sexual partner. ``Autosomes'' are the ``normal'' chromosomes 
which are not the X or Y sex chromosome.
Bottom part: Age structure of females and males in the above population.
From \cite{Biecek}.
}
\end{figure}

\begin{figure}[hbt!]
\begin{center}
\includegraphics[angle=-90,scale=0.35]{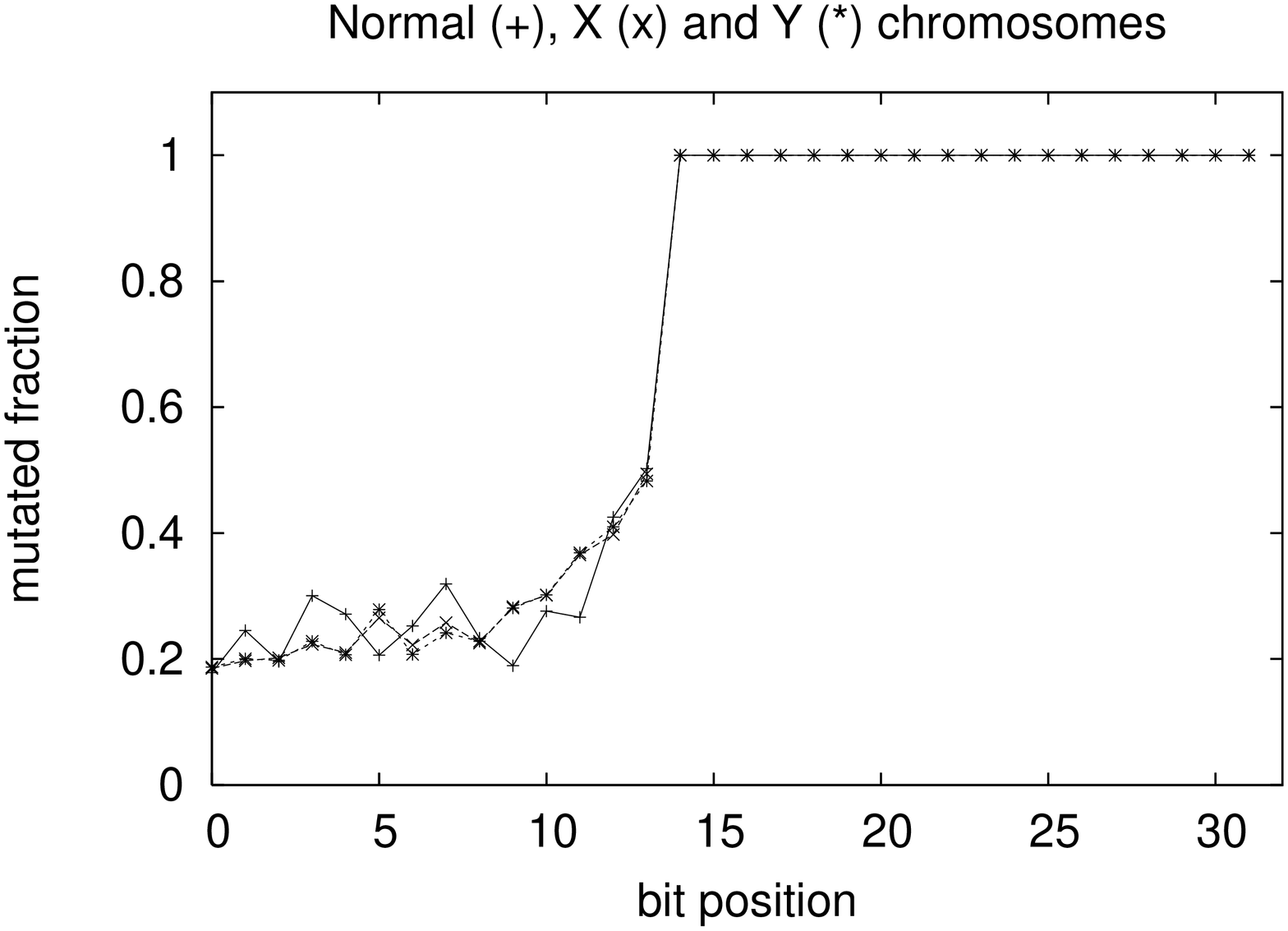}
\includegraphics[angle=-90,scale=0.35]{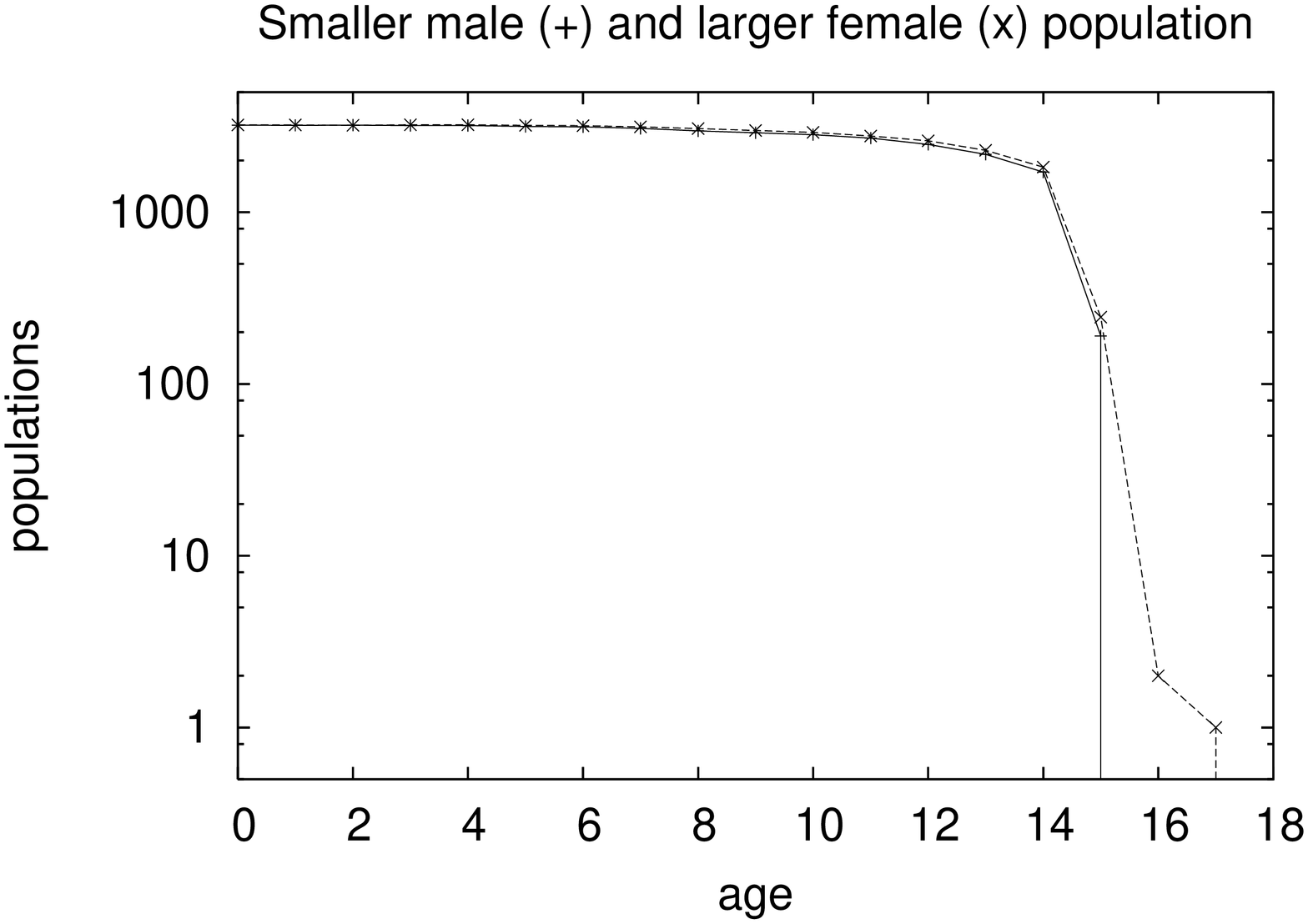}
\end{center}
\caption{As preceding figure but for faithful men. The Y chromosome now behaves
like the other chromosomes.
From \cite{Biecek}.
}
\end{figure}

\begin{figure}[hbt]
\begin{center}
\includegraphics[angle=-90,scale=0.45]{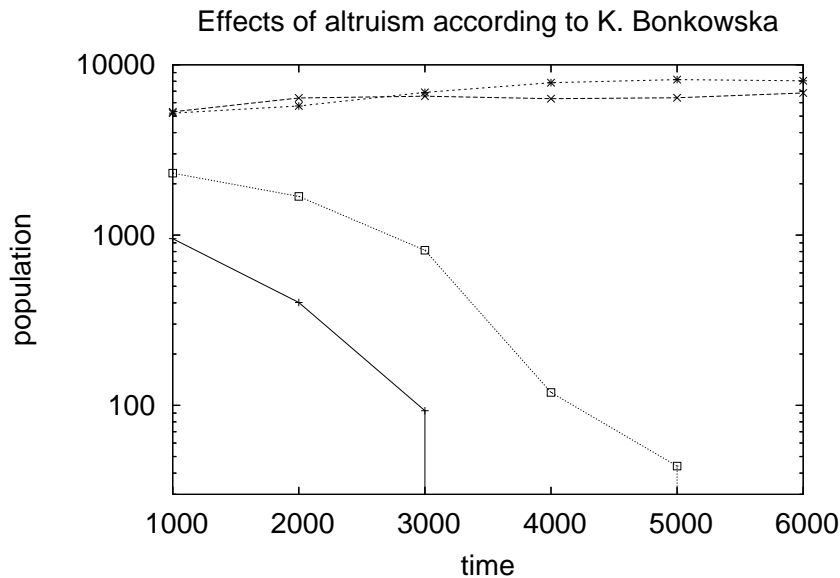}
\end{center}
\caption{Growth and decay for the four altruism choices: Women sacrificed (+),
women saved by men (x), men sacrificed (stars), men saved by women (squares).
}
\end{figure}

To model the necessity of the male altruism, Bo\'nkowska \cite{thesis}
introduced into the Penna model the random death for specific sex at
specific age (random in this case means that individuals at a declared age
and sex could be killed with a declared probability independently of their
genetic status). These individuals, in some instances could be saved by
other individuals but then, the savior has to die.
She described the parameters of four populations:

\begin{itemize}

\item females at age 10 are randomly killed with probability 0.2 and nobody
tries to save them,

\item females at age 10 are killed randomly with probability 0.2 and males at
age 90 try to save them with probability 0.5. If a man succeeds in saving
the oppressed young woman - he dies,

\item males at age 10 are randomly killed with probability 0.2 and nobody tries
to save them,

\item males at age 10 are killed randomly with probability 0.2 and females at
age 90 try to save them with probability 0.5. If a female succeeds in saving
the stupid man - she dies.

\end{itemize}

Other parameters of simulations are the same for all four populations: the
length of the bitstring is $L=128$, minimum reproduction age 80, menopause 110,
birth rate 0.2. Simulations were started with the same size of each
population. All four populations were placed in one environment where they
have to compete. The results are shown in Fig.9. The results are so
pronounced that they need a special attention. The worst situation is when
women die randomly without any help. This decreases the population
reproduction potential and the population looses in competition. The other
loosing strategy is when women try to save men. It is much better if the
oppressed women may count on the male altruism; then the population is almost as
good as when a part of men are eliminated randomly. The evolution of
altruism is much more complicated because it depends on the age at which the
men sacrifice their life (not shown here). It is better if they are younger,
that is why one can hardly find an old general in the front line.

All these examples show how crucial is the role of male genomes in the human
genetic pool evolution.  But there is one puzzle. It is assumed that the
mutation rate per one cell generation (per genome replication) is
constant \cite{drake}.
It has been estimated that the number of cell divisions from zygote to the
sexual maturity of woman (production of the egg in the oogenesis in the female
germ line) is 24 according to \cite{vogel} 
and it is roughly the same for all eggs produced by a woman during her whole
reproduction period. The corresponding process of sperm production
(spermatogenesis) is quite different. The number of cell divisions during
the spermatogenesis for 20 years old men is estimated as 150 and
increases by 23 each year \cite{hurst}. The resulting higher mutation numbers
in male gametes was one reason to warn against the too high cost of sex for 
females \cite{redfield}. 

One can expect that the
mutation rate per male gamete should grow very fast linearly with age. The
experimental data are inconsistent; for some genes the sex bias in the
mutation rate is observed while for some other genes is not. This is also
contradictory with the results shown in the above sections which clearly
indicate that the role of the older men is very important in keeping the
longer lifespan of humans of both sexes. That is why it could be
biologically legitimate to assume that there is the above-mentioned gamete 
preselection during or before the fertilization.

\bigskip

It would be completely wrong to assert that we are against women. We are very 
grateful for them to have reverted the 2002 final of the male-football world 
cup. The senile author thanks the senior editor for movie education.
We thank COST P10 for supporting the visit of DS.

\section{Appendix 1: Penna ageing model}

The Penna model of mutation accumulation describes well the biological ageing, 
with a mortality increasing exponentially with adult age, and has been reviewed 
in detail in \cite{books}; the present description is adapted from 
\cite{zawierta}. The sexual population is composed of a changing number 
$N$ of individuals, each represented by its 
diploid ``genome'' corresponding to two bit-strings (haplotypes) $L=64$ bits 
long. At an age $x$, only the first $x$ bits are active. 
Bits set to 0 or 1 correspond to the correct (wild) or defective genes, 
respectively. Only if both genes are set for 1 at the same position on the 
bit-strings, the effect of the ,,locus'' (bit position) on the individual
(having reached at least the age corresponding to the bit position)
is bad and the individual dies because of the genetic death. Thus all defects 
are recessive and one single active mutation kills, and the older the
individual gets, the more bit positions are checked for possible defects. 

Females try to give $B=2$ births if they have reached at least the minimum 
reproduction age $R=40$ and found a male partner, randomly selected from the 
whole population, who is also old enough. Each attempt succeeds with probability
$N(t)/N_{\max}$, which is the Verhulst factor due to limited food and space.
To give a birth, the female genome is replicated, during the replication one new
mutation is introduced for each copy of bit-string 
into a randomly chosen locus. If the bit at the locus is 0 it is replaced by 
$1$, if it is 1 it stays 1 (there are no reversions). The two copies of 
bit-strings recombine with probability $r$ at a randomly chosen point just by 
exchanging the corresponding arms (crossover). After these processes, each of 
the two new bit-strings corresponds to a gamete. A randomly chosen female gamete
is joined with another gamete produced in the same way by the male 
partner. The pair of gametes corresponds to the newborn's diploid genome. 
The newborn's sex is established with equal probability to be male or female. 
(Some parameters were selected differently in \cite{bonkowska,thesis,Biecek}.)

Sex is complicated; the asexual version has only one bit-string, and no 
recombination. Fortran programs are published by Moss de Oliveira et al
\cite{books}. 

\section{Appendix 2: Future demographic problems}

As the reader may have noticed, these authors became old. Over most of the 
world, life expectancies have increased and birth rates have fallen. Who 
will feed us when we become even older?

Some people claim that demographic predictions of the future are just 
propaganda to help the government to reduce old-age pensions. Similarly one
could claim that warnings of man-made global warming is just propaganda for
nuclear power plants; and if their is really more sunshine, we just buy more 
sunglasses. Of course, any prediction of future ratios of old to young people
is just theory and could be wrong. Similarly, the assertion that all people
die at some time, or that the sun will rise tomorrow, is merely theory. It
becomes invalid if an immortality gene is found and activatead, or if Bruce 
Willis fails to prevent the impact of a huge meteorite on Earth.  

Future age distributions have been simulated with the Penna model in 
\cite{demography}. Such simulations take into account heredity and slow
changes in the human genome. If we restrict ourselves to the 21st century,
the hereditary correlations between mother and daughter do not have enough time
to change, and instead of an agent-based simulation \cite{demography} of 
individuals one may
evaluate the changes in probability distributions as in \cite{redfield}, which
is much simpler \cite{stauffer,martinsret,bomsdorf,zekri,sumour,poland}. 

(The difference between the proper way \cite{demography} and the approximation
by distributions \cite{stauffer,martins,zekri,bomsdorf,sumour,poland}
can be explained by a simple example. Let the present 
population consist of two groups such that on average, women have 1.9 children
in one and 2.1 children in the other group, and let us assume that this
difference is propagated culturally or via mitochondrial DNA from mother to
daughter, without any change. Then, after many generations most of the 
population will belong to 
the group with the higher number of births, but that effect is ignored if only 
one age distribution is simulated. However, until the year 2050 we do not have 
that many generations to get this domination of one group, and a simulation
with an average  number of 2.0 children per women remains sufficient.)

Also, babies who die in their first years neither require schooling nor pensions
and may be neglected in a simulation of the number of people in retirement age
to the number of people in working age. The birth rates and the mortalities
put into such a simulation are thus those for adults: How many people reach
the working age (20 years, for example), and how many reach retirement age.
Finally, the number of people older than 110 years, when mortality might reach
a plateau, is still very small (one in a million for West-Central Europe)
\cite{robine}. Thus we can use the Gompertz law, Fig.6, that mortality 
increases exponentially with age. Either the medical progress since two 
centuries leads to a rectangularisation of survival probabilities, as a result
of which 
asymptotically we will run Marathon races in two hours at the age of 102, and
die of old age within the following year. Or, alternatively, the survival curves
since about 1970 shift to older age without changing anymore their shape 
\cite{tuljapurkar}. Simple Fortran programs using this alternative are 
published in \cite{sumour} and \cite{books}b.

The results are roughly the same in the various types of simulated countries 
\cite{europe}: If
the number of children per women (fecundity; often misleadingly called the total
fertility rate) sinks far below the replacement level near 2.1, problems
appear decades later when the ratio of retired to working people becomes very
high. When that will happen depends on the time development of this number of 
births, which is near 1.3 in Germany \cite{stauffer,bomsdorf} since a third of a
century (similarly in Poland \cite{poland} since a shorter time), increased 
from 1.7 to 1.9 in France from 1995 to 2005, is close to the 
replacement level in Algeria \cite{zekri} and is still much higher in the 
Palestinian territories \cite{sumour}. 

Possible remedies are immigration \cite{stauffer}, increases in retirement
ages \cite{martins}, more work by women \cite{sumour} or an increased birth 
rate \cite{poland}: Romania nearly
doubled the number of births per women, for a short time around 1968. (These
last choices allow to make women responsible for our problems.) Only
people were simulated, not money, since the buying power of pensions may be 
changed by law or by inflation.


\begin{thebibliography}{99}

\bibitem{schneider} J.J. Schneider, S. Kirkpatrick, Stochastic Optimization,
Springer, Berlin, Heidelberg, New York, 2006.

\bibitem{cebpek} S. Cebrat, A. P\c {e}kalski, Eur. Phys. J. B 11, 687 (1999).

\bibitem{whysex} J. S. Sa Martins and S. Moss de Oliveira, Int. J. Mod. Phys.
C 9, 421 (1998).

\bibitem{books}  S. Moss de Oliveira, P.M.C. de Oliveira, D. Stauffer:
{\it Evolution, Money, War and Computers}, Teubner, Stuttgart and Leipzig 1999;
D. Stauffer, S. Moss de Oliveira, P.M.C. de Oliveira, J.S. S\'a Martins,
\textit {Biology, Sociology, Geology by Computational Physicists}. Amsterdam:
Elsevier 2006.

\bibitem{howard} R.S. Howard, C.M. Lively, Nature 367, 554 (1994).

\bibitem{martins} J.S. S\'a Martins, Phys. Rev. E 61, 2212 (2000).

\bibitem{redfield} R.J. Redfield, Nature  369, 145 (1994).

\bibitem{zorzenon} D. Stauffer, P.M.C. de Oliveira, S. Moss de Oliveira,
R.M. Zorzenon dos Santos, Physica A  231, 504 (1996).

\bibitem{anais} D. Stauffer, P.M.C. de Oliveira, S. Moss de Oliveira, J.S. 
S\'a Martins, T.J.P. Penna, Anais Acad. Bras. Cienc. 73, 15 (2001). 

\bibitem{penna} T.J.P. Penna: J. Stat. Phys. 78, 1629 (1995). 

\bibitem{weidlich} W. Weidlich, {\it Sociodynamics: A Systematic Approach to Mathematical Modelling in the Social Sciences} (Gordon and Breach, London, 2000).

\bibitem{martinsstauffer} J.S. S\'a Martins, D. Stauffer, Physica A 294, 191
(2001).

\bibitem{zawierta} M. Zawierta, P. Biecek, W. Waga and S. Cebrat, The role of 
intragenomic recombination rate in the evolution of population's genetic pool. 
Theory in Biosciences 125, 124. (2007).

\bibitem{bonkowska}
K. Bo\'nkowska, M. Kula, S. Cebrat, D. Stauffer, Int. J. Mod. Phys. C 18 (8), 
in press (2007).

\bibitem{pekalski}
A. P\c {e}kalski,  Int. J. Mod. Phys. C 18 (10), in press (2007).

\bibitem{pmco}
P.M.C. de Oliveira, S. Moss de Oliveira, D. Stauffer, S. Cebrat, A. 
P\c {e}kalski, preprint: Does sex induce a phase transition? arXiv:0710.1357
submitted to Eur. Phys. J. B..

\bibitem{childcare} S. Moss de Oliveira, A.T. Bernardes, J.S. S\'a Martins, 
Eur. Phys. J. B7, 501 (1999).

\bibitem{lahdenpera} M. Lahdenper\"a, V. Lummaa, S. Helle, M. Tremblay, A.F. 
Russell, Nature 428, 178 (2004).

\bibitem{gamete} S. Cebrat, D. Stauffer, Int. J. Mod. Phys. C 19, in press 
(2008), e-print 0709.2420 at arXiv.org \quad .

\bibitem{schneiderold}
J. Schneider, S. Cebrat, D. Stauffer, 
International Journal of Modern Physics C 9, 721 (1998)

\bibitem{encybio}
D. Stauffer,  The Penna Model of Biological Aging, to be published in
Bioinformatics and Biology Insights (2007). 

\bibitem{blickpunkt}
M. Dinges, Blickpunkt: Der Mann 4,21 (2006) (in German language).

\bibitem{robine} J.-M. Robine, J.W. Vaupel, Exp. Gerontology 36, 915
(2001). See also K. Suematsu, M. Kohno, J. Theor. Biol. 201, 231 (1999) and
for opposing conclusions 
N.S. Gavrilova, L.A. Gavrilov 2005. Search for Predictors of Exceptional Human
Longevity. In: ``Living to 100 and Beyond'' Monograph. The Society of
Actuaries, Schaumburg, Illinois, USA, pp. 1-49.

\bibitem{onody}
M.P.Lobo, R.N. Onody, Eur. Phys. J. B  45, 533 (2005) and Physica A 361, 239 
(2006). 

\bibitem{klotz} D. Stauffer, T. Klotz: The Aging Male, 4, 95 (2001);
D. Stauffer, Theory in Biosc. 120, 87 (2001).

\bibitem{drake}
J.W. Drake, B. Charlesworth, D. Charlesworth, J.F. Crow, Genet. 148, 1667 
(1998).

\bibitem{vogel} F. Vogel, A.G. Motulsky, Human
Genetics, Problems and Approaches (3rd edn), Springer, Berlin-Heidelberg 1997. 

\bibitem{hurst} L.D. Hurst, H. Ellegren, Trends in Genetics 14, 446 (1998).

\bibitem{Niewczas} E. Niewczas, S. Cebrat, D. Stauffer
Theory Biosci. 119, 122 (2000).

\bibitem{Kurdziel} E. Niewczas, A. Kurdziel, S. Cebrat, 
Int. J. Mod. Phys. C. 11, 775 (2000).

\bibitem{Copp} A.J. Copp, Trends Genet., 11, 87 (1995).

\bibitem{Hassold} T.J. Hassold, Trends Genet., 2, 87 (1986).

\bibitem{Biecek} P. Biecek, S. Cebrat, Why Y (chromosome) shrinks, to
be published (2008)

\bibitem{thesis} K. Bo\'nkowska PhD thesis, Wroc{\l}aw University 2008.

\bibitem{demography} K. Bo\'nkowska, S. Szymczak and C. Cebrat,
Int. J. Mod. Phys. C 17, 1477 (2006).

\bibitem{stauffer} D. Stauffer, Exp. Gerontology 37, 1131 (2002).

\bibitem{martinsret} J.S. S\'a Martins, D. Stauffer, Ingenierias
(Univ.Nuevo Leon, Mexico) 7, 35 (Jan-Mar issue) (2004).

\bibitem{bomsdorf} E. Bomsdorf, Exp. Gerontology 39, 159 (2004).

\bibitem{zekri} L. Zekri, D. Stauffer, AIP Conference Proceedings 779,
69 (2005).

\bibitem{sumour} M. A. Sumour, A. H. El-Astal, M. M. Shabat and M. A. Radwan,
Int. J. Mod. Phys. C 18, issue 11 (2007).

\bibitem{poland} D. Stauffer, e-print arXiv:0709.4389 ???.

\bibitem{tuljapurkar} R.D. Edwards, S. Tuljapurkar.  Population and
Development Review 31, 645 (2005), in particular Fig.6 there.

\bibitem{europe} Page 5 in: http://www.staat-modern.de/Anlage/original\_1120864/Finland.pdf


\end{thebibliography}
\end{document}